\documentclass[12pt,fleqn]{article}
\usepackage{amsmath}

\begin{document}

\title{\bf Enlarged spectral problems\\
and nonintegrability}

\author{{\sc Sergei Sakovich}\\[12pt] {\small Institute of Physics,
National Academy of Sciences,}\\[-4pt] {\small 220072 Minsk,
Belarus. E-mail: saks@tut.by}}

\date{}

\maketitle

\begin{abstract}
The method of obtaining new integrable coupled equations through
enlarging spectral problems of known integrable equations, which
was recently proposed by W.-X.~Ma, can produce nonintegrable
systems as well. This phenomenon is demonstrated and explained by
the example of the enlarged spectral problem of the
Korteweg--de~Vries equation.
\end{abstract}

\section{Introduction}

Recently, Ma \cite{Ma1} proposed a method of obtaining new
integrable systems of coupled equations through enlarging spectral
problems of known integrable equations, and successfully applied
it to the AKNS hierarchy. More recently, Ma \cite{Ma2} applied
this method to the hierarchy of vector AKNS equations, also
successfully. The new hierarchies were proven to be integrable in
the sense of possessing recursion operators and infinite sets of
commuting symmetries.

The method of Ma consists in using the enlarged first-order Lax
pair
\begin{equation} \label{e1}
\Psi_x = X \Psi , \qquad \Psi_t = T \Psi
\end{equation}
with the square matrices $X$ and $T$ of the block form
\begin{equation} \label{e2}
X =
\begin{pmatrix}
U & A \\ 0 & 0
\end{pmatrix}
, \qquad T =
\begin{pmatrix}
V & B \\ 0 & 0
\end{pmatrix}
,
\end{equation}
where the square submatrices $U$ and $V$ correspond to the Lax
pair $\Phi_x = U \Phi$ and $\Phi_t = V \Phi$ of a known integrable
equation, and the submatrices $A$ and $B$ contain additional
dependent variables. The compatibility condition
\begin{equation} \label{e3}
D_t X = D_x T - [ X , T ]
\end{equation}
of the linear equations \eqref{e1} is equivalent to
\begin{align}
D_t U & = D_x V - [ U , V ] \label{e4} ,\\ D_t A & = D_x B - U B +
V A \label{e5} ,
\end{align}
where $D_t$ and $D_x$ stand for the total derivatives, and the
square brackets denote the commutator. Therefore the condition
\eqref{e3} is a zero-curvature representation (ZCR) of a coupled
system which consists of the initial integrable equation
determined by \eqref{e4} and an additional equation determined by
\eqref{e5}.

In the present paper, we show that --- and explain why --- the
method of Ma can produce nonintegrable coupled equations. We
enlarge the spectral problem of the Korteweg--de~Vries (KdV)
equation and study the evolutionary systems possessing the ZCR
\eqref{e3} with the matrix $X$ of the form
\begin{equation} \label{e6}
X =
\begin{pmatrix}
0 & 1 & p \\ u + \lambda & 0 & q \\ 0 & 0 & 0
\end{pmatrix}
,
\end{equation}
where $u$, $p$ and $q$ are functions of $x$ and $t$, and $\lambda$
is the spectral parameter. Throughout the paper, we set
$T_{31}=T_{32}=T_{33}=0$, thus considering matrices $T$ of the
block form \eqref{e2} only, in compliance with the method of Ma.

In section~\ref{s2}, we choose the matrix $T$ to be of a definite
special form, such that the ZCR \eqref{e3} determines a
one-parameter class of coupled KdV-type equations, and show by
means of the Painlev\'{e} analysis that at least ten systems of
that class cannot be expected to be integrable by the inverse
scattering transform technique. In section~\ref{s3}, we obtain the
complete class of systems of local evolution equations possessing
Lax pairs \eqref{e1} with $X$ given by \eqref{e6} and traceless
matrices $T$ of the block form \eqref{e2}, and observe that most
of those systems cannot be integrable in any reasonable sense. In
section~\ref{s4}, we explain this strangeness of the studied
spectral problem by the existence of a gauge transformation which
merges two of the three dependent variables of the matrix $X$
\eqref{e6} into one new dependent variable. Section~\ref{s5}
contains concluding remarks.

\section{Singularity analysis} \label{s2}

Let us choose the matrix $T$ to be of the following special form:
\begin{equation} \label{e7}
T =
\begin{pmatrix}
u_x & - 2 u + 4 \lambda & T_{13} \\ u_{xx} - 2 u^2 + 2 \lambda u +
4 \lambda^2 & - u_x & T_{23} \\ 0 & 0 & 0
\end{pmatrix}
\end{equation}
with
\begin{equation} \label{e8}
\begin{split}
T_{13} & = 4 p_{xx} + 4 q_x - 2 u p + 4 \lambda p , \\ T_{23} & =
4 q_{xx} - p u_x + k u p_x - 2 u q + 4 \lambda p_x + 4 \lambda q ,
\end{split}
\end{equation}
where $k$ is a parameter. Then the ZCR \eqref{e3} with the
matrices $X$ \eqref{e6} and $T$ \eqref{e7}--\eqref{e8} determines
the following one-parameter class of triangular systems of coupled
KdV-type equations:
\begin{equation} \label{e9}
\begin{split}
u_t & = u_{xxx} - 6 u u_x , \\ p_t & = 4 p_{xxx} - ( k + 2 ) u p_x
, \\ q_t & = 4 q_{xxx} - 6 u q_x - 3 q u_x + ( k - 4 ) u p_{xx} +
( k - 1 ) u_x p_x .
\end{split}
\end{equation}

In order to show that the method of Ma can produce nonintegrable
coupled equations, it is sufficient to find any value of the
parameter $k$ such that the system \eqref{e9} fails to pass some
reliable test for integrability. We choose the Painlev\'{e} test
in its version for partial differential equations \cite{WTC,JKM}.
More precisely, we use the presence of nondominant logarithmic
singularities in solutions of a nonlinear system as an undoubted
sign of its nonintegrability by the inverse scattering transform
technique \cite{Tab,AC}.

Trying to represent the singular behavior of solutions of the
system \eqref{e9} by the expansions
\begin{equation} \label{e10}
u = \sum_{i=0}^{\infty} u_i (t) \phi^{i + \alpha}, \qquad p =
\sum_{i=0}^{\infty} p_i (t) \phi^{i + \beta}, \qquad q =
\sum_{i=0}^{\infty} q_i (t) \phi^{i + \gamma}
\end{equation}
with $\partial_x \phi (x,t) = 1$, we find that $\alpha = - 2$,
$\beta = - b$ and $\gamma = - b - 1$, and that the positions of
resonances are $i = -1, 0, 4, 6, b, b, b+2, b+4, 2b+3$, where $b$
is related to the parameter $k$ by
\begin{equation} \label{e11}
k = 2 b^2 + 6 b + 2 .
\end{equation}
Non-integer exponents of the dominant behavior of solutions and
non-integer positions of resonances not necessarily imply
nonintegrability: for example, the integrable Harry Dym equation
possesses such analytic properties. Since our aim is to find any
undoubtedly nonintegrable case of \eqref{e9}, we continue the
singularity analysis for integer values of $b$ only, in order to
discover the incompatibility of recursion relations at resonances,
which is the strongest indication of nonintegrability.

Substituting the expansions \eqref{e10} to the system \eqref{e9},
we get the following recursion relations for their coefficients
$u_i , p_i , q_i$:
\begin{equation} \label{rr}
\begin{split}
& (n-2) (n-3) (n-4) u_n - 3 (n-4) \sum_{i=0}^n u_i u_{n-i} \\ &
\qquad - (n-4) \phi_t u_{n-2} - u_{n-3,t} = 0 , \\[16pt] & 4 (n-b)
(n-b-1) (n-b-2) p_n \\ & \qquad - (k+2) \sum_{i=0}^n (n-b-i) u_i
p_{n-i} \\ & \qquad - (n-b-2) \phi_t p_{n-2} - p_{n-3,t} = 0 ,
\\[16pt] & 4 (n-b-1) (n-b-2) (n-b-3) q_n \\ & \qquad - \sum_{i=0}^n
(6n -6b -3i -12) u_i q_{n-i} \\ & \qquad + \sum_{i=0}^n (n-b-i)
\bigl( (k-4) (n-b) -3k +3i +6 \bigr) u_i p_{n-i} \\ & \qquad -
(n-b-3) \phi_t q_{n-2} - q_{n-3,t} = 0 ,
\end{split}
\end{equation}
where $ n = 0, 1, 2, \dotsc $, and $u_i = p_i = q_i = 0$ for $i <
0$. Let us check the compatibility of these recursion relations
for $b=1,2,3,\dotsc$. In the case of $b = 1$, when $k = 10$ due to
\eqref{e11}, the nontrivial compatibility condition $p_0 \,
\phi_{tt} = 0$ appears for \eqref{rr} at the double resonance
$i=5$. This means that we have to modify the expansions
\eqref{e10} by introducing additional logarithmic terms, starting
from the position $i = 5$ in this case. Moreover, in at least nine
further cases, with $b = 2, 3, \dotsc , 10$, the recursion
relations \eqref{rr} turn out to be incompatible as well:
nontrivial compatibility conditions appear there in the position
$i = b$. For example, we get the condition $p_0 \phi_t =0$ at the
double resonance $i=2$ when $b=2$, the condition $p_{0,t} =0$ at
the double resonance $i=3$ when $b=3$, the condition $ 160 u_4 p_0
- 3 p_0 \phi_t^2 =0$ at the triple resonance $i=4$ when $b=4$, the
condition $1675 p_0 \phi_{tt} - 52 p_{0,t} \phi_t =0$ at the
double resonance $i=5$ when $b=5$, the condition $ 166053888 u_6
p_0 +44876160 u_4 p_0 \phi_t - 446631 p_0 \phi_t^3 - 572 p_{0,tt}
=0$ at the triple resonance $i=6$ when $b=6$, and so on.

Consequently, at least ten systems of the class \eqref{e9}
--- namely, those with $k=10,22,38,58,82,110,142,178,218,262$ ---
cannot be expected to be integrable by the inverse scattering
transform technique, due to the presence of nondominant
logarithmic singularities in their solutions. We have found that
the Lax pair \eqref{e1} with the matrices $X$ \eqref{e6} and $T$
\eqref{e7}--\eqref{e8} behaves ill: it represents nonintegrable
systems, like the so-called weak Lax pairs do, which contain no
essential parameter (see, e.g., \S~5.1 in \cite{CC} about the
nonintegrable Dodd--Fordy equation). On the other hand, this Lax
pair \eqref{e1} with \eqref{e6}--\eqref{e8} should be called
strong, because it contains the essential parameter $\lambda$. In
the next two sections, we study how a strong Lax pair can be weak.

\section{Continual class} \label{s3}

Let us find the complete class of systems of local evolution
equations
\begin{equation} \label{e12}
\begin{split}
u_t & = f ( x , t , u , p , q , \dotsc , u_{x \dotsc x} , p_{x
\dotsc x} , q_{x \dotsc x} ) , \\ p_t & = g ( x , t , u , p , q ,
\dotsc , u_{x \dotsc x} , p_{x \dotsc x} , q_{x \dotsc x} ) , \\
q_t & = h ( x , t , u , p , q , \dotsc , u_{x \dotsc x} , p_{x
\dotsc x} , q_{x \dotsc x} )
\end{split}
\end{equation}
which possess the ZCR \eqref{e3} with the predetermined matrix $X$
\eqref{e6} and any traceless $(2+1) \times (2+1)$-dimensional
matrix $T$ of the block form \eqref{e2}. It is convenient to solve
this problem by the cyclic basis method \cite{Sak1,Sak2}.

The characteristic form of the ZCR \eqref{e3} is
\begin{equation} \label{e13}
f C_u + g C_p + h C_q = \nabla T ,
\end{equation}
where $C_u = \partial X / \partial u$, $C_p = \partial X /
\partial p$ and $C_q = \partial X / \partial q$, and the operator
$\nabla$ is defined by $\nabla Y = D_x Y - [ X , Y ]$ for any $3
\times 3$ matrix $Y$. The linearly independent matrices $C_u$,
$C_p$, $C_q$, $\nabla C_u$ and $\nabla^2 C_u$ constitute the
cyclic basis with the closure equations
\begin{equation} \label{e14}
\begin{split}
& \nabla C_p = - ( u + \lambda ) C_q , \qquad \nabla C_q = - C_p ,
\\[2pt] & \nabla^3 C_u = 2 u_x C_u - 3 ( p_x + q ) C_p + ( p_{xx} + q_x
) C_q + 4 ( u + \lambda ) \nabla C_u .
\end{split}
\end{equation}
This five-dimensional cyclic basis is sufficient for decomposing
any $3 \times 3$ matrix $T$ with $T_{31} = T_{32} = T_{33} =
T_{11} + T_{22} = 0$ over it, as follows:
\begin{equation} \label{e15}
T = a_1 C_u + a_2 C_p + a_3 C_q + a_4 \nabla C_u + a_5 \nabla^2
C_u .
\end{equation}
Then, from \eqref{e13}, \eqref{e15} and \eqref{e14}, we find that
\begin{equation} \label{e16}
a_1 = D_x^2 a_5 - 4 ( u + \lambda ) a_5 , \qquad a_4 = - D_x a_5 ,
\end{equation}
and that
\begin{align}
f & = D_x^3 a_5 - 4 ( u + \lambda ) D_x a_5 - 2 u_x a_5 ,
\label{e17} \\ g & = D_x a_2 - a_3 - 3 ( p_x + q ) a_5 ,
\label{e18} \\ h & = - ( u + \lambda ) a_2 + D_x a_3 + ( p_{xx} +
q_x ) a_5 . \label{e19}
\end{align}
The relations \eqref{e17}--\eqref{e19} determine all the sought
systems \eqref{e12} in terms of all functions $a_2 , a_3 , a_5$ of
$\lambda , x , t , u , p , q , \dotsc , u_{x \dotsc x} , p_{x
\dotsc x} , q_{x \dotsc x}$ such that the conditions $\partial f /
\partial \lambda = \partial g / \partial \lambda = \partial h /
\partial \lambda = 0$ are satisfied.

It is helpful to rewrite the relations \eqref{e17}--\eqref{e19} in
the form
\begin{equation} \label{e20}
F = \bigl( K - \lambda L \bigr) A ,
\end{equation}
where
\begin{align}
F & =
\begin{pmatrix} f \\ g \\ h
\end{pmatrix}
, \qquad A =
\begin{pmatrix} a_2 \\ a_3 \\ a_5
\end{pmatrix}
, \label{FA} \\[6pt] K & =
\begin{pmatrix}
0 & 0 & D_x^3 - 4 u D_x - 2 u_x \\ D_x & -1 & - 3 p_x - 3 q \\ - u
& D_x & p_{xx} + q_x
\end{pmatrix}
, \label{e21} \\[6pt] L & =
\begin{pmatrix}
0 & 0 & 4 D_x \\ 0 & 0 & 0 \\ 1 & 0 & 0
\end{pmatrix}
. \label{e22}
\end{align}
If the cyclic basis method, being applied to a certain spectral
problem, leads to a relation of the form \eqref{e20} and the
operator $L^{-1}$ exists there, then one can immediately conclude
that the class of local evolutionary systems associated with that
spectral problem is a discrete hierarchy with the recursion
operator $R = K L^{-1}$ \cite{Sak2,Sak3}. Indeed, by using the
expansion $A = A_0 + \lambda A_1 + \lambda^2 A_2 + \dotsb$ (if $A$
is not analytic at $\lambda = 0$, a shift of $\lambda$ is
required, with no effect on $L$) one finds from \eqref{e20} and
$\partial F
/
\partial \lambda = 0$ that $F = K A_0$ and $K A_{i+1} = L A_i$ ($i
= 0, 1, 2, \dotsc$), which show that $F = K L^{-1} F'$ is the
right-hand side of a represented system if and only if $F'$ is. For example, one can easily prove that the
operator $L^{-1}$ exists in the case of the enlarged AKNS spectral
problem studied in \cite{Ma1}, and that the operator $R = K
L^{-1}$ is in that case exactly the recursion operator obtained in
\cite{Ma1}. However, the present case of the enlarged KdV
equation's spectral problem is manifestly different: we see from
\eqref{e22} that the operator $L^{-1}$ does not exist.

In order to overcome this observed degeneracy of the relations
\eqref{e17}--\eqref{e19}, we use \eqref{e18} as a definition of
$a_3$,
\begin{equation} \label{e23}
a_3 = D_x a_2 - 3 ( p_x + q ) a_5 - g ,
\end{equation}
and then obtain from \eqref{e17} and \eqref{e19} the following:
\begin{equation} \label{e24}
\begin{pmatrix}
f \\ s
\end{pmatrix}
=
\Bigl( M - \lambda N \Bigr)
\begin{pmatrix}
a_2 \\ a_5
\end{pmatrix}
,
\end{equation}
where
\begin{align}
M & =
\begin{pmatrix}
0 & D_x^3 - 4 u D_x - 2 u_x \\ D_x^2 - u & - 3 v D_x - 2 v_x
\end{pmatrix}
, \label{e25} \\[6pt] N & =
\begin{pmatrix}
0 & 4 D_x \\ 1 & 0
\end{pmatrix}
, \label{e26} \\[6pt] s & = D_x g + h , \qquad v = p_x + q .
\label{27}
\end{align}
Since the operator $N^{-1}$ exists for $N$ \eqref{e26}, we can
immediately conclude from \eqref{e24} that the pairs of functions
$f$ and $s$ satisfying the conditions $\partial f /
\partial \lambda = \partial s / \partial \lambda = 0$ constitute
a discrete hierarchy generated by the recursion operator $R = 4 M
N^{-1}$ (the factor $4$ is taken for simplicity):
\begin{equation} \label{e28}
R =
\begin{pmatrix}
D_x^2 - 4 u - 2 u_x D_x^{-1} & 0 \\[2pt] - 3 v - 2 v_x D_x^{-1} &
4 D_x^2 - 4 u
\end{pmatrix}
.
\end{equation}
Note, however, that the function $g$ remains arbitrary because no
restriction appeared on it.

The operator $R$ \eqref{e28} is the G\"{u}rses--Karasu recursion
operator of the new integrable coupled KdV system
\begin{equation} \label{e29}
u_t = u_{xxx} - 6 u u_x , \qquad v_t = 4 v_{xxx} - 6 u v_x - 3 v
u_x
\end{equation}
discovered in \cite{GK}. It was shown in \cite{Sak4} that the
system \eqref{e29} passes the Painlev\'{e} test well and possesses
a $(3 \times 3)$-dimensional Lax pair (equivalent to a special
case of the enlarged KdV equation's Lax pair, actually).

Thus, we have found the following continual class of evolutionary
systems \eqref{e12} possessing the enlarged Lax pair
\eqref{e1}--\eqref{e2} with $X$ determined by \eqref{e6}:
\begin{equation} \label{e30}
u_t = f , \qquad p_t = g , \qquad q_t = s - D_x g ,
\end{equation}
where
\begin{equation} \label{e31}
\begin{pmatrix}
f \\ s
\end{pmatrix}
= R^n
\begin{pmatrix}
0 \\ 0
\end{pmatrix}
, \qquad n = 0 , 1 , 2 , \dots ,
\end{equation}
$R$ is given by \eqref{e28} with $v = p_x + q$, the quantity
$D_x^{-1} 0$ is understood as any function of $t$, and $g$ is an
arbitrary function of $x, t, u, p, q$ and $x$-derivatives of $u,
p, q$ of any maximal order not related to $n$. Though the matrix
$X$ \eqref{e6} contains the essential parameter $\lambda$ which
cannot be removed by a gauge transformation of the ZCR, as one can
see from the gauge-invariant coefficients of the closure equations
\eqref{e14}, the obtained class of systems \eqref{e30} is a
continual class containing the arbitrary function $g$, not a
discrete hierarchy that could be expected in a strong Lax pair
case. We have seen in section~\ref{s2} that some systems of the
class \eqref{e30}, due to their analytic properties, cannot be
expected to be integrable through the inverse scattering
transform. Moreover, we believe that most of systems \eqref{e30}
are nonintegrable in any reasonable sense. Indeed, the evident
reduction $u = 0$, $q = - p_x$, made in \eqref{e30}, leads to the
class of all scalar evolution equations $p_t = \tilde{g} ( x, t,
p, p_x , \dotsc , p_{x \dotsc x} )$, but there is no sense to
speak about integrability of a generic evolution equation.

\section{Gauge transformation} \label{s4}

The observed strangeness of the studied Lax pair can be explained
by the existence of a gauge transformation
\begin{equation} \label{e32}
\begin{split}
\Psi' & = G \Psi , \qquad \det G \neq 0 , \\ X' & = G X G^{-1} + (
D_x G ) G^{-1} , \\ T' & = G T G^{-1} + ( D_t G ) G^{-1}
\end{split}
\end{equation}
which merges two of the three dependent variables of the matrix
$X$ \eqref{e6} into one new dependent variable. Indeed, the
transformation \eqref{e32} with
\begin{equation} \label{e33}
G =
\begin{pmatrix}
1 & 0 & 0 \\ 0 & 1 & p \\ 0 & 0 & 1
\end{pmatrix}
\end{equation}
changes the matrix $X$ \eqref{e6} into
\begin{equation} \label{e34}
X' =
\begin{pmatrix}
0 & 1 & 0 \\ u + \lambda & 0 & v \\ 0 & 0 & 0
\end{pmatrix}
,
\end{equation}
where $v = p_x + q$.

Let us find what the matrix $T$ of the studied Lax pair is changed
into by this gauge transformation. It follows from \eqref{e15},
\eqref{e16} and \eqref{e23} that the matrix $T$ corresponding to
the class \eqref{e30} has the form
\begin{equation} \label{iT}
T =
\begin{pmatrix}
D_x a_5 & -2 a_5 & a_2 -2 p a_5 \\ D_x^2 a_5 - 2 ( u + \lambda )
a_5 & - D_x a_5 & T_{23} \\ 0 & 0 & 0
\end{pmatrix}
,
\end{equation}
where
\begin{equation} \label{T23}
T_{23} = D_x a_2 - p D_x a_5 - 2 ( p_x + q ) a_5 - g .
\end{equation}
Applying the gauge transformation \eqref{e32} with $G$ \eqref{e33}
to \eqref{iT}--\eqref{T23}, we get
\begin{equation} \label{tT}
T' =
\begin{pmatrix}
D_x a_5 & -2 a_5 & a_2 \\ D_x^2 a_5 - 2 ( u + \lambda ) a_5 & -
D_x a_5 & D_x a_2 - 2 v a_5 + p_t - g \\ 0 & 0 & 0
\end{pmatrix}
.
\end{equation}
It is easy to see from \eqref{e24}--\eqref{e26}, \eqref{e28} and
\eqref{e31} that $a_2$ and $a_5$ are functions of $\lambda, u, v,
u_x, v_x, \dotsc, u_{x \dotsc x}, v_{x \dotsc x}$ only. Therefore
the transformed matrix $T'$ \eqref{tT} contains $p$ and $q$ in the
form of the new dependent variable $v = p_x + q$ everywhere,
except for the expression $p_t - g$ in $T'_{23}$, but $p_t - g =
0$ on all solutions of any represented system \eqref{e30}.

Consequently, up to the gauge equivalence and the equivalence on
solutions \cite{Sak1}, the studied `strong-and-weak' enlarged Lax
pair effectively contains only one additional dependent variable
but not two.

\section{Conclusion} \label{s5}

In the present paper, we studied the method of W.-X.~Ma of
obtaining new integrable systems of coupled equations through
enlarging spectral problems of known integrable equations. We
enlarged the spectral problem of the KdV equation and found that
the method of Ma can produce nonintegrable systems as well. We
explained this strangeness of the enlarged KdV equation's spectral
problem by the existence of a gauge transformation which merges
two of the three dependent variables of the enlarged spectral
problem into one new dependent variable. Let us remind that for
the first time a phenomenon of this kind, that nonintegrable
systems can possess a ZCR containing an essential parameter, was
observed and explained through gauge transformations in
\cite{Sak1}.

\end{document}